\documentclass[10pt]{article}
\usepackage[T1]{fontenc}
\usepackage[utf8]{inputenc}
\usepackage{amsmath,amssymb,amsthm,mathtools}
\usepackage{preprint}
\usepackage{graphicx,booktabs,tabularx,multirow,array,makecell,adjustbox}
\usepackage{caption,subcaption}
\usepackage{placeins,stfloats,balance,enumitem}
\usepackage[numbers,sort&compress]{natbib}
\usepackage[hidelinks]{hyperref}
\usepackage{microtype}
\usepackage{xurl,etoolbox}
\providecommand{\Description}[2][]{}
\makeatletter
\def\verbatim@font{\normalfont\ttfamily\small}
\makeatother

\let\arxivOriginalIncludegraphics\includegraphics
\renewcommand{\includegraphics}[2][]{%
\ifstrequal{#2}{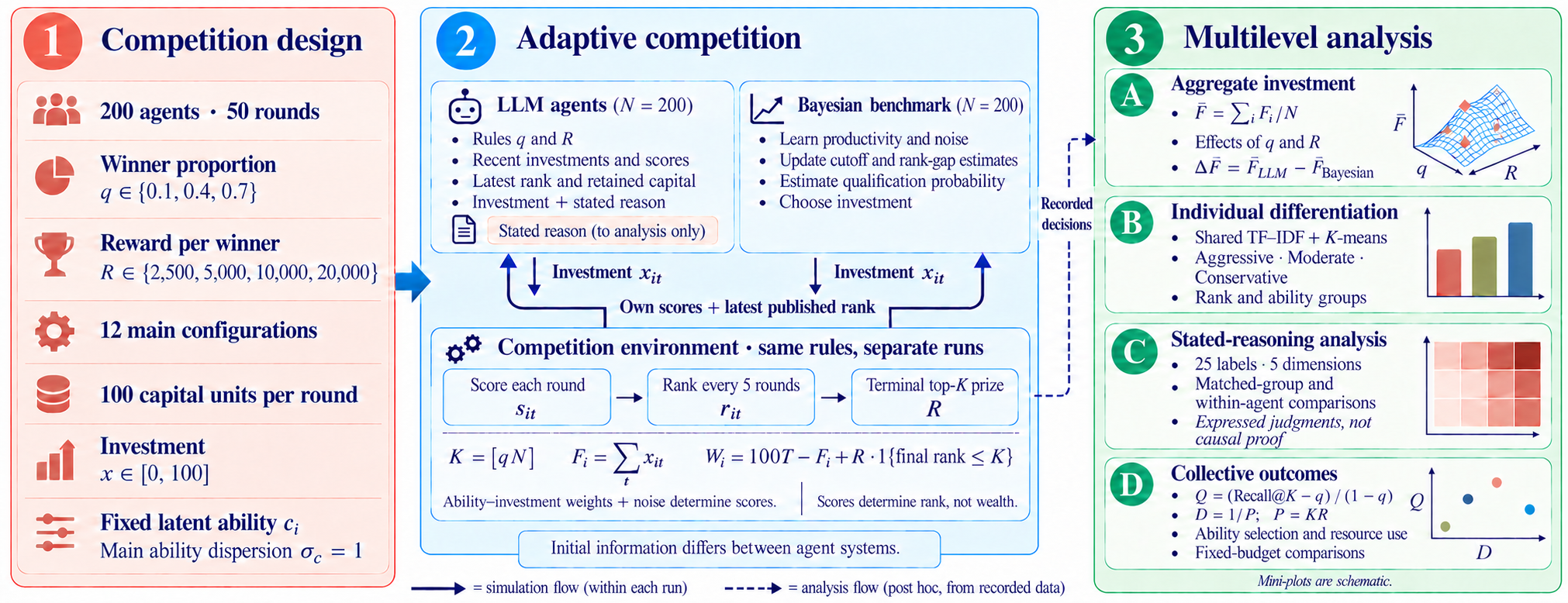}{\arxivOriginalIncludegraphics[#1,width=0.94\textwidth,height=0.34\textheight,keepaspectratio]{#2}}{\ifstrequal{#2}{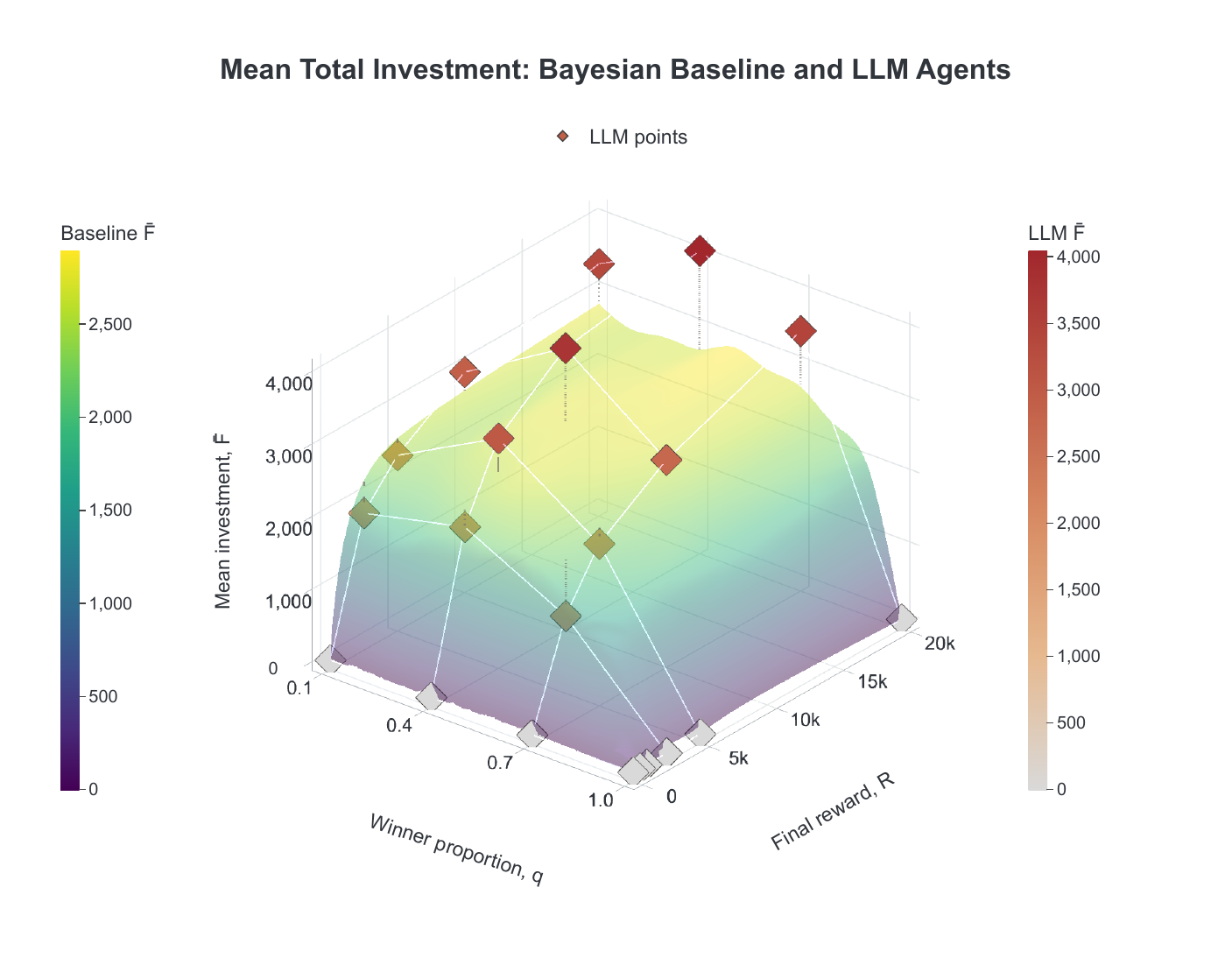}{\arxivOriginalIncludegraphics[#1,width=0.64\textwidth,height=0.38\textheight,keepaspectratio]{#2}}{\ifstrequal{#2}{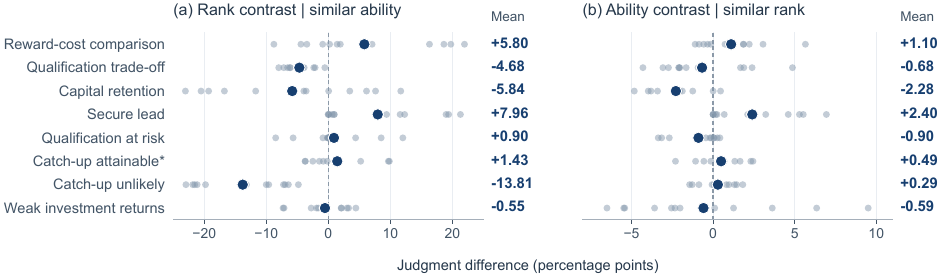}{\arxivOriginalIncludegraphics[#1,width=0.88\textwidth,height=0.3\textheight,keepaspectratio]{#2}}{\ifstrequal{#2}{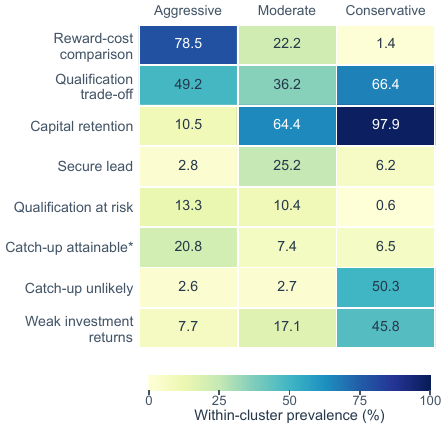}{\arxivOriginalIncludegraphics[#1,width=0.66\textwidth,height=0.36\textheight,keepaspectratio]{#2}}{\ifstrequal{#2}{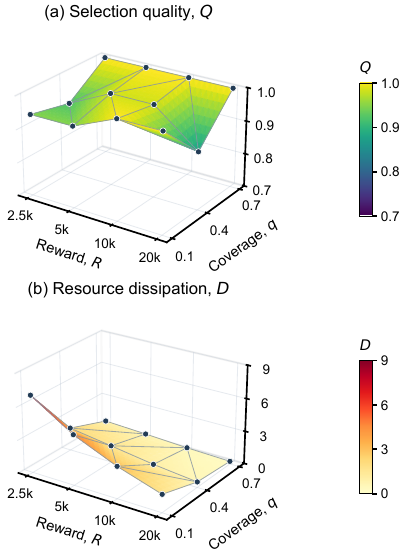}{\arxivQdRow}{\ifstrequal{#2}{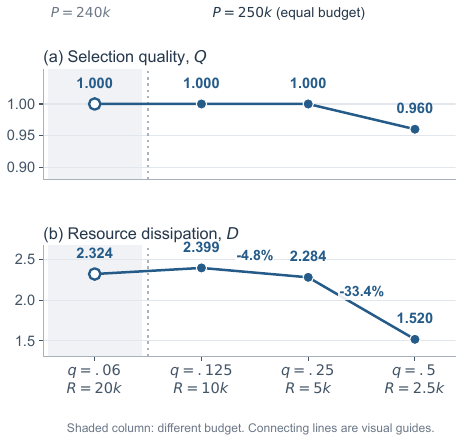}{\arxivOriginalIncludegraphics[#1,width=0.66\textwidth,height=0.36\textheight,keepaspectratio]{#2}}{\ifstrequal{#2}{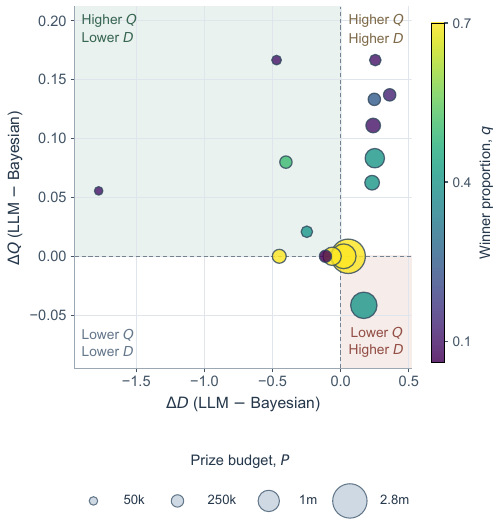}{\arxivOriginalIncludegraphics[#1,width=0.6\textwidth,height=0.4\textheight,keepaspectratio]{#2}}{\ifstrequal{#2}{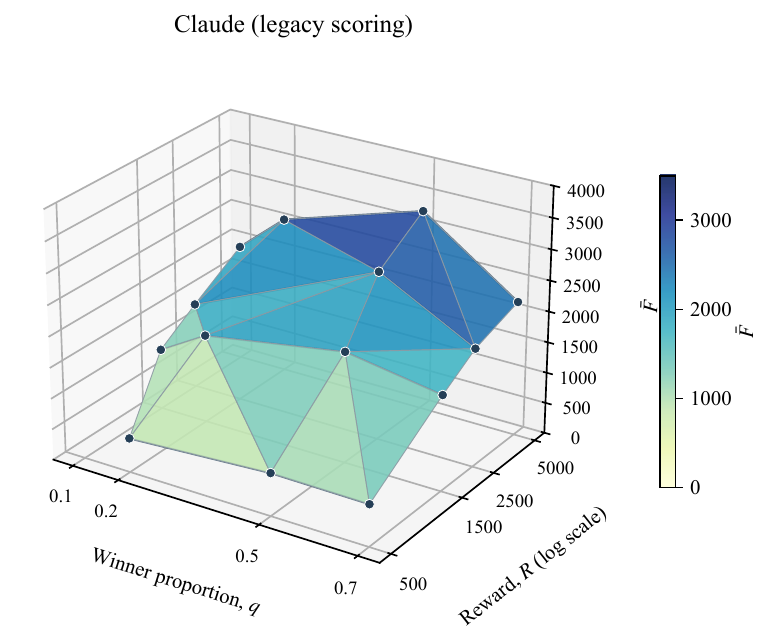}{\arxivOriginalIncludegraphics[#1,width=0.58\textwidth,height=0.36\textheight,keepaspectratio]{#2}}{\ifstrequal{#2}{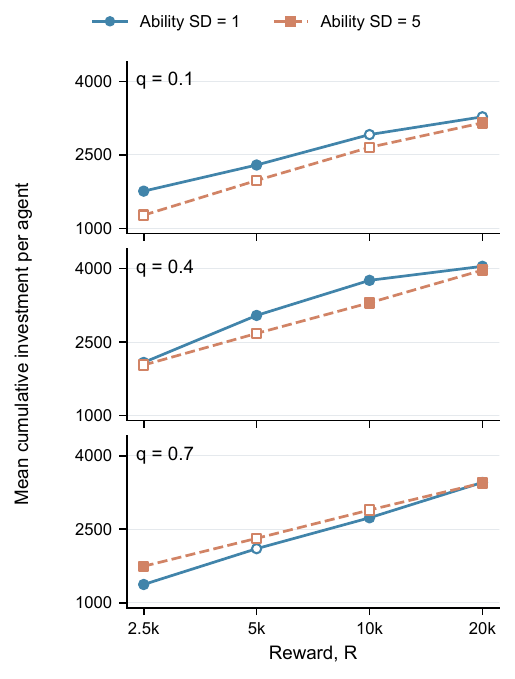}{\arxivAbilityRow}{\arxivOriginalIncludegraphics[#1]{#2}}}}}}}}}}%
}

\newcommand{\arxivQdRow}{\scalebox{1}[1.1764706]{\makebox[\textwidth]{%
\arxivOriginalIncludegraphics[viewport=0 132.4 194.327 264.788,clip,width=.49\textwidth]{sections/collective_qd_single_column_z07.pdf}\hfill
\arxivOriginalIncludegraphics[viewport=0 0 194.327 132.4,clip,width=.49\textwidth]{sections/collective_qd_single_column_z07.pdf}}}}
\newcommand{\arxivAbilityPanel}[2]{\begin{minipage}[t]{.32\textwidth}\centering
\arxivOriginalIncludegraphics[viewport=28 #1 246.888 #2,clip,width=\linewidth]{sections/ability_dispersion_compact.pdf}\\[-2pt]
\arxivOriginalIncludegraphics[viewport=28 0 246.888 31,clip,width=\linewidth]{sections/ability_dispersion_compact.pdf}\end{minipage}}
\newcommand{\arxivAbilityRow}{\begin{minipage}{\textwidth}\centering
\arxivOriginalIncludegraphics[viewport=0 303 246.888 325.467,clip,width=.50\textwidth]{sections/ability_dispersion_compact.pdf}\\[2pt]
{\small Mean cumulative investment per agent (50 rounds)}\\[3pt]
\arxivAbilityPanel{211}{299}\hfill\arxivAbilityPanel{121}{209}\hfill\arxivAbilityPanel{31}{119}
\end{minipage}}
\AtBeginEnvironment{tabular}{\renewcommand{\arraystretch}{1.15}}
\date{}

\title{The Price of a Chance in Artificial Societies}
\author{Shuqi Zhao\textsuperscript{1}\quad Boyu Wang\textsuperscript{4}\quad Shuangyue Niu\textsuperscript{3}\quad Chenhao Zhang\textsuperscript{2}\quad Zhanzhan Zhao\textsuperscript{1,2,*}}
\preprintaffiliations{
\textsuperscript{1}School of Humanities and Social Science, The Chinese University of Hong Kong, Shenzhen, Shenzhen, China\\[2pt]
\textsuperscript{2}School of Artificial Intelligence, The Chinese University of Hong Kong, Shenzhen, Shenzhen, China\\[2pt]
\textsuperscript{3}School of Data Science, The Chinese University of Hong Kong, Shenzhen, Shenzhen, China\\[2pt]
\textsuperscript{4}Department of Mathematics, The Ohio State University, Columbus, USA\\[2pt]
\textsuperscript{*}Corresponding author: Zhanzhan Zhao (\href{mailto:zhanzhanzhao@cuhk.edu.cn}{zhanzhanzhao@cuhk.edu.cn})}
\begin{document}
\maketitle
\begin{abstract}
Rapid advances in large language models (LLMs) have enabled autonomous agents to interact, coordinate, and pursue individual goals in artificial societies. However, our understanding of how reward structures shape their competitive behavior and collective outcomes remains incomplete. We simulate 200 LLM agents competing over 50 rounds to maximize individual wealth. We define overcompetition and lying flat as population investment above and below a Bayesian adaptive baseline, respectively. At the lowest tested reward, populations lie flat; at the two highest, they overcompete, most strongly at intermediate winner coverage. The investment committed to pursuing an uncertain reward constitutes the price of a chance. Agents near the qualification cutoff can invest heavily yet receive no prize, finishing poorer than nonwinners who invest less. Agents' stated rationales link investment decisions to perceived qualification prospects and expected returns. At the collective level, greater investment does not consistently improve ability selection. With a fixed total prize budget, offering smaller prizes to more winners can reduce investment while preserving selection quality relative to the corresponding top-ability target. Further broadening reward coverage yields additional savings but reduces selection quality. These findings show how reward design shapes overcompetition and lying flat, distributes the costs of pursuing opportunity, and influences what competition collectively achieves.
\end{abstract}
\vspace{12pt}
\section{Introduction}
\label{sec:introduction}
\begingroup
\setlength{\emergencystretch}{2em}

Multi-agent systems built from large language models (LLMs) are being developed for collaborative problem solving, including software engineering and supply-chain decision support~\cite{wu2024autogen,hong2024metagpt}. Research also examines agents that negotiate on behalf of buyers and sellers~\cite{zhu2025automated}. When agents represent different users or organizational interests, their objectives may be only partially aligned. Where rewards depend on relative performance, competition becomes a problem of system design: reward rules can change how much agents invest, who benefits, and how effectively the system uses resources. Understanding these responses matters for multi-agent systems whose participants pursue individual success.

Competition also offers a scientific opportunity to investigate how incentives become differentiated investment decisions. LLM agents can interpret natural-language rules and performance histories, reason about potential returns, and express judgments about their competitive prospects~\cite{park2023generative,mei2024turing,akata2025games}. These capacities allow investment to be studied alongside psychologically interpretable assessments of security, threatened qualification, and the attainability of catching up, without prescribing aggressive or conservative responses in advance. Published studies have shown that LLM populations can develop social conventions and collective biases~\cite{ashery2025conventions}, and that networked interactions can generate polarization~\cite{piao2026polarization}. These findings motivate examining how competitive institutions organize context-dependent judgments and actions into population outcomes, providing a setting for formulating hypotheses about persistence, withdrawal, and the collective consequences of competition.

Rank-based rewards make this problem especially sharp. Only a specified share of participants receives a prize, but everyone chooses investment before knowing whether they will qualify. We call this committed investment \emph{the price of a chance}. The price is paid whether or not an agent ultimately qualifies. A final ranking records winners but reveals little about nonwinners' investment. Changing reward magnitude or coverage can alter both aggregate investment and its distribution. The question is whether that investment improves selection, and who pays for the process.

Tournament theory explains how rank-based rewards can incentivize effort~\cite{lazear1981tournaments}, while contest research examines how prize allocation, information, and participant heterogeneity affect investment and outcomes~\cite{dechenaux2015survey,cason2020winner,boosey2020information}. Agent-based models connect individual adaptation to population outcomes~\cite{macy2002actors}, including escalation and withdrawal under prescribed strategies and updating rules~\cite{wang2021replicator,wang2022modeling,chaocheng2023involution}. These foundations make it possible to ask how LLM agents interpret incentive structures in practice, and how differences in their responses connect individual investment with collective selection.

We study 200 LLM agents competing over 50 rounds to maximize individual final wealth. Agents invest part of each round's endowment, retain the remainder, and receive periodic rank feedback. They interact indirectly through relative scores and shared reward rules, without direct communication. An environment-assigned ability parameter determines performance productivity, rather than the underlying model's intelligence. Terminal rewards go to a specified proportion of the population. We vary reward magnitude $R$ and winner proportion $q$, and compare investment with a Bayesian adaptive heuristic. At the population level, we define \emph{overcompetition} as investment above this baseline and \emph{lying flat} as investment below it. Agents receive no assigned aggressive or conservative personalities, allowing differentiated responses to be examined under common instructions.

Three research questions connect the price of a chance to reward design, individual differentiation, and collective outcomes:
\begin{enumerate}
    \item \textbf{RQ1. When does reward design produce overcompetition or lying flat?} We examine how reward magnitude and coverage change investment relative to the baseline and how competitive strategies evolve.
    \item \textbf{RQ2. Who pays the price of a chance?} We examine how investment and stated judgments about continued competition differ across ranks and abilities.
    \item \textbf{RQ3. What does this price achieve for the collective?} We examine whether greater investment improves ability selection and whether reallocating an equal prize budget can reduce investment while preserving selection quality.
\end{enumerate}

The results connect three consequences of reward design. Rewards can move a population from lying flat to overcompetition; they can concentrate high investment among agents that ultimately receive no reward; and they can mobilize resources without consistently improving ability selection. Equal-budget comparisons further show that investment can be reduced while preserving selection quality relative to each allocation's target, although further savings introduce a trade-off. These findings connect competitive escalation and withdrawal with the unequal price of pursuing a chance and the collective returns that investment achieves.
\endgroup

\section{Methodology}
\label{sec:methodology}
\begingroup
\setlength{\emergencystretch}{1.5em}
\label{sec:simulation-framework}
We develop a simulation of repeated rank-based competition in which LLM agents and a Bayesian adaptive benchmark operate in separate populations under matched competition rules and intrinsic abilities. Figure~\ref{fig:study-workflow} summarizes the environment, decision--feedback process, and multilevel analysis linking individual decisions to collective outcomes.

\begin{figure*}[!t]
  \centering
  \includegraphics[width=\textwidth]{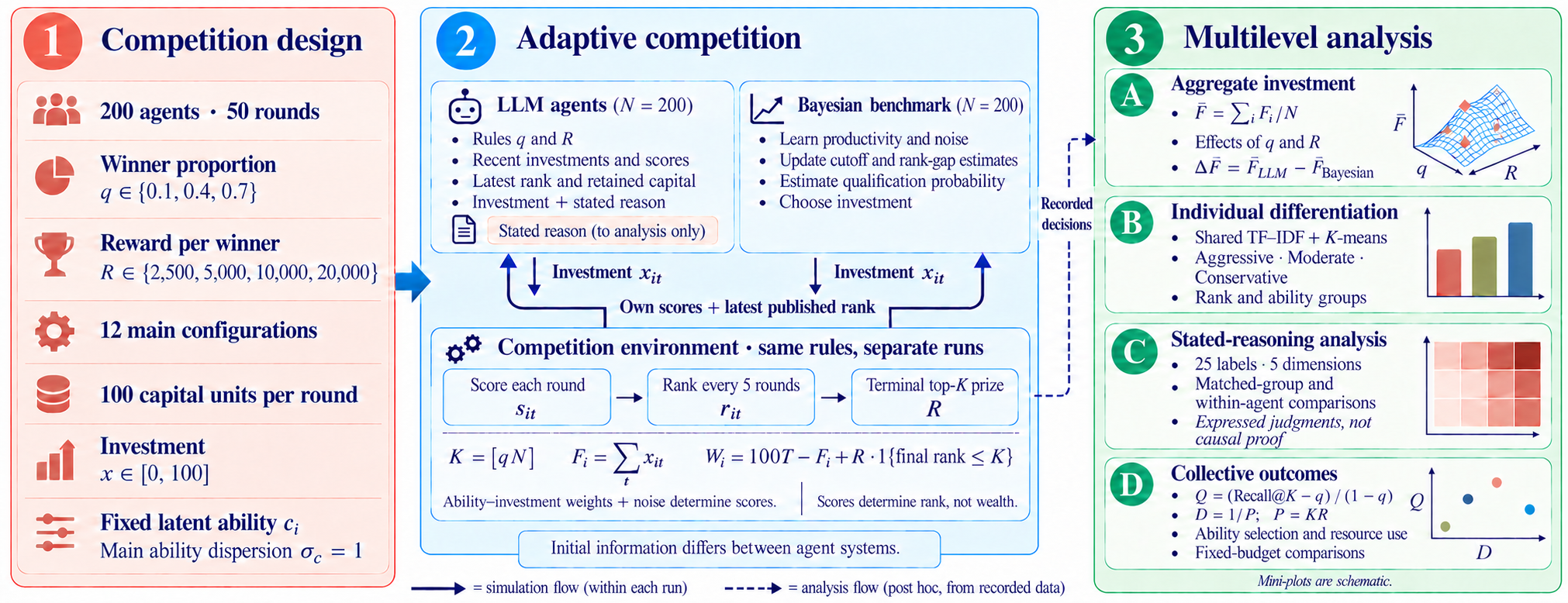}
  \caption{Study workflow linking competition design, adaptive decisions, and multilevel analysis. LLM and Bayesian populations run independently under matched rules and abilities. Solid arrows denote simulation flows and dashed arrows post hoc analysis; mini-plots are schematic.}
  \Description{Three modules depict competition design, distinct LLM and Bayesian decision systems with a common depiction of environment rules and separate feedback paths, and four analyses illustrated by schematic surface, bar, heatmap, and scatter plots.}
  \label{fig:study-workflow}
\end{figure*}

\subsection{Competitive Environment}
\label{sec:competitive-environment}
Each agent $i$ has a fixed latent ability $c_i$. In each round $t$, it receives $E$ capital units and chooses investment $x_{it}\in[0,E]$. Uninvested capital contributes to final wealth but cannot be reinvested in subsequent rounds. Ability and investment jointly determine the agent's share of a fixed scoring pool $M$, with Gaussian noise added and negative scores truncated:
\begin{equation}
s_{it}=\max\!\left\{0,\,
M\frac{\exp(c_i x_{it}/\kappa)}{\sum_{j=1}^{N}\exp(c_j x_{jt}/\kappa)}+\epsilon_{it}\right\},
\quad \epsilon_{it}\sim\mathcal N(0,\sigma_\epsilon^2).
\label{eq:competition-score}
\end{equation}
Here, $\kappa$ controls the sensitivity of score allocation to ability--\allowbreak{}investment differences. The scoring pool $M$ is fixed independently of realized investment and the prize budget. Because ability is signed, additional investment raises an agent's deterministic score share when $c_i>0$ and lowers it when $c_i<0$, holding others' investments fixed.

Agents receive their own scores after each round. Rankings based on cumulative average scores are published every five rounds. At horizon $T$, the top $K=\lceil qN\rceil$ agents receive reward $R$. Final wealth is
\begin{equation}
W_i=ET-\sum_{t=1}^{T}x_{it}+R\mathbf 1\{r_{iT}\le K\},
\label{eq:terminal-wealth}
\end{equation}
where $r_{iT}$ denotes final rank, with smaller values indicating better performance. Scores determine reward eligibility but are not directly added to wealth.

\subsection{Adaptive Decision Systems}
\label{sec:adaptive-decisions}
\paragraph{LLM agents.}
Each agent is instructed to maximize final wealth. Its decision context contains the competition rules, current round, retained capital, cumulative investment, up to five recent rounds of its own investments and scores, and the latest published rank and cumulative average score. No rank is available before the first publication; between publications, ranking signals remain unchanged while round-level histories update. True ability affects performance but is not supplied numerically. The prompt guides agents to assess qualification prospects and investment returns, and requests an investment and a stated reason in structured JSON. The environment processes decisions, computes scores, and updates the next decision context. Appendix~\ref{app:llm-prompt} provides the prompt and feedback sequence.

\paragraph{Bayesian adaptive benchmark.}
The benchmark estimates local investment productivity, performance noise, rank-related score deficits, and investment near the qualification cutoff from recent observations. For each candidate investment, it approximates terminal qualification probability, averages over sampled parameter settings, and maximizes estimated reward minus current expenditure. Its updates combine Bayesian-style precision updates with smoothed empirical estimates; the approximation and decision equations appear in Appendix~\ref{app:benchmark}.

The benchmark provides an adaptive investment reference rather than an equilibrium solution or normative efficiency bound. Its runner initializes a private noisy rank before public feedback, whereas LLM agents initially receive no rank. Comparisons therefore concern the implemented systems under their respective information conditions.

\subsection{Analysis and Measures}
\label{sec:analysis-measures}
\paragraph{Aggregate investment.}
We calculate individual cumulative investment $F_i=\sum_t x_{it}$, population investment $I=\sum_i F_i$, and mean cumulative investment $\bar F=I/N$ from saved round-level decisions. The signed gap
\begin{equation}
\Delta\bar F=\bar F_{\mathrm{LLM}}-\bar F_{\mathrm{Bayesian}}
\label{eq:investment-gap}
\end{equation}
indicates greater or lower investment relative to the benchmark.

\paragraph{Individual differentiation.}
We pool stated rationales at Rounds 5, 15, 25, 35, and 45 across the main configurations, then apply K-means to a shared TF--IDF representation. Clusters are named aggressive, moderate, and conservative in descending order of their associated mean investment; these names describe observed clusters rather than predefined strategies. PCA visualizes the shared representation. Rank comparisons distinguish the reward zone (RZ, top $K$) and consecutive thirds of remaining ranks (NM, MID, LOW), using within-band proportions. Fixed ability groups comprise the upper 25\%, middle 50\%, and lower 25\%. Clustering and grouping details appear in Appendix~\ref{app:clustering}.

\paragraph{Stated reasoning.}
We code rationales using 25 overlapping labels across five dimensions: reward--cost reasoning, perceived competitive position, investment efficacy and ability, action intent, and others' behavior. Strategy comparisons, matched rank/ability comparisons, and within-agent analyses examine differences in expressed judgments. Results report the comparison samples; Appendix~\ref{app:reasoning} provides the codebook, matching procedures, and aggregation protocol.

\paragraph{Collective outcomes.}
Let $\mathcal A_K$ contain the top-$K$ agents by intrinsic ability and $\mathcal W_K$ the final winners. We define
\begin{equation}
\mathrm{Recall}@K=\frac{|\mathcal A_K\cap\mathcal W_K|}{K},\qquad
Q=\frac{\mathrm{Recall}@K-q}{1-q},\qquad
D=\frac{I}{P},\quad P=KR.
\label{eq:collective-measures}
\end{equation}
Selection quality $Q=1$ indicates perfect recovery and $Q=0$ chance-level recovery; $Q$ is undefined at $q=1$. Changing $q$ changes the target set size. Lower dissipation $D$ indicates less investment per unit of prize budget. We interpret it alongside absolute investment and use equal-budget comparisons to assess selection--expenditure trade-offs. These measures evaluate ability selection and resource use, rather than comprehensive social welfare.

\subsection{Experimental Settings}
\label{sec:experimental-design}
The main simulations use \texttt{gpt-5.6-sol}. Each simulation contains $N=200$ agents over $T=50$ rounds, with $E=100$ capital units per round. Abilities are sampled from a normal distribution with mean $0.5$ and standard deviation $\sigma_c=1$, then clipped to $[-100,100]$. Scoring parameters are $M=20{,}000$, $\kappa=20$, and $\sigma_\epsilon=5$. We report one LLM run per configuration and benchmark seed~0, with matched intrinsic abilities.

The main experiment crosses $q\in\{0.1,\allowbreak 0.4,\allowbreak 0.7\}$ with $R\in\{2{,}500,\allowbreak 5{,}000,\allowbreak 10{,}000,\allowbreak 20{,}000\}$, yielding 12 configurations. Budget-controlled comparisons use $(q,R)$ pairs $(0.06,20{,}000)$, $(0.125,10{,}000)$, $(0.25,5{,}000)$, and $(0.5,2{,}500)$. The latter three share $P=250{,}000$; the first has $P=240{,}000$ and serves as a nearby-budget reference rather than a strictly equal-budget configuration.

\par\endgroup

\section{Results}
\subsection{RQ1. Reward Design Shapes Overcompetition and Lying Flat}
\label{sec:aggregate-investment}
\label{sec:differentiation}
\begin{figure}[!tbp]
\centering
\includegraphics[width=\linewidth]{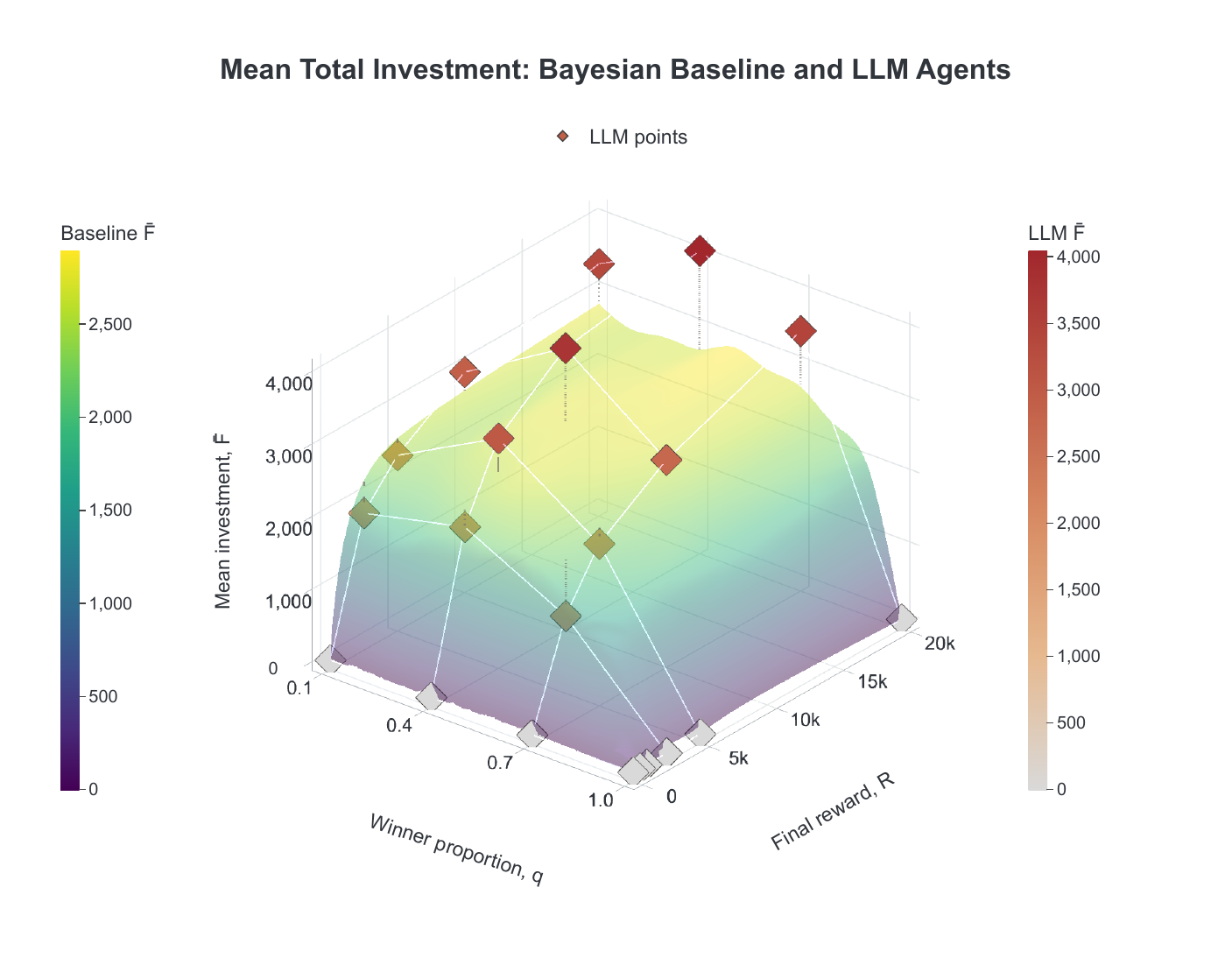}
\caption{Average cumulative investment per agent across reward magnitude $R$ and winner proportion $q$. The surface represents the Bayesian adaptive benchmark, and diamonds mark LLM observations. Connecting lines and the surface are visual guides.}
\Description{Comparison of LLM and Bayesian average cumulative investment across rewards and winner proportions. LLM investment peaks at the tested intermediate winner proportion and increases with rewards.}
\label{fig:aggregate-investment}
\end{figure}

\begin{figure*}[!t]
\centering
\begin{minipage}[t]{0.49\textwidth}\centering\vspace{0pt}
\scalebox{1}[0.85]{\includegraphics[width=\linewidth]{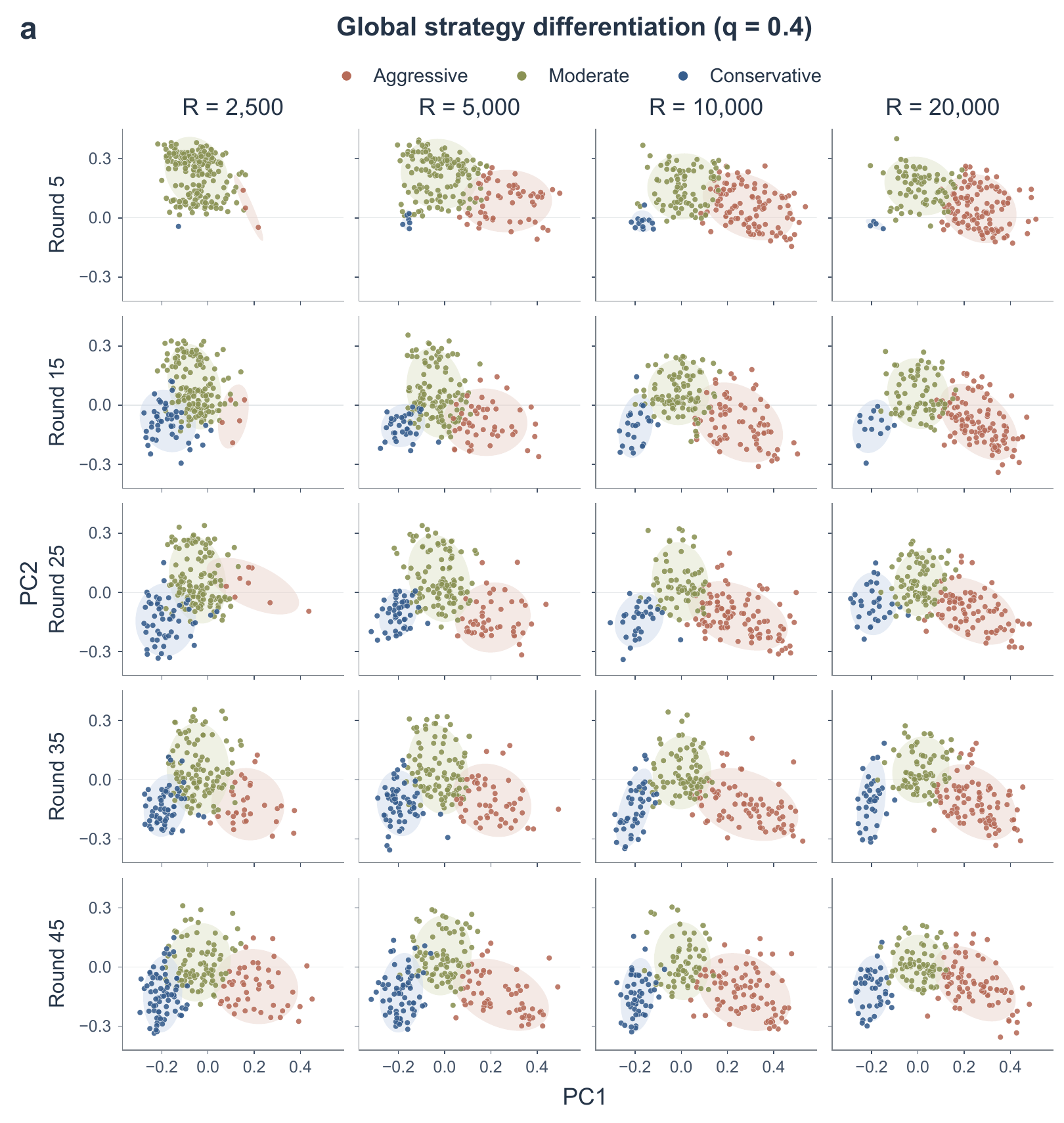}}\par\smallskip
\small (a) Fixed $q=0.4$, varying $R$.
\end{minipage}\hfill
\begin{minipage}[t]{0.49\textwidth}\centering\vspace{0pt}
\scalebox{1}[0.85]{\includegraphics[width=\linewidth]{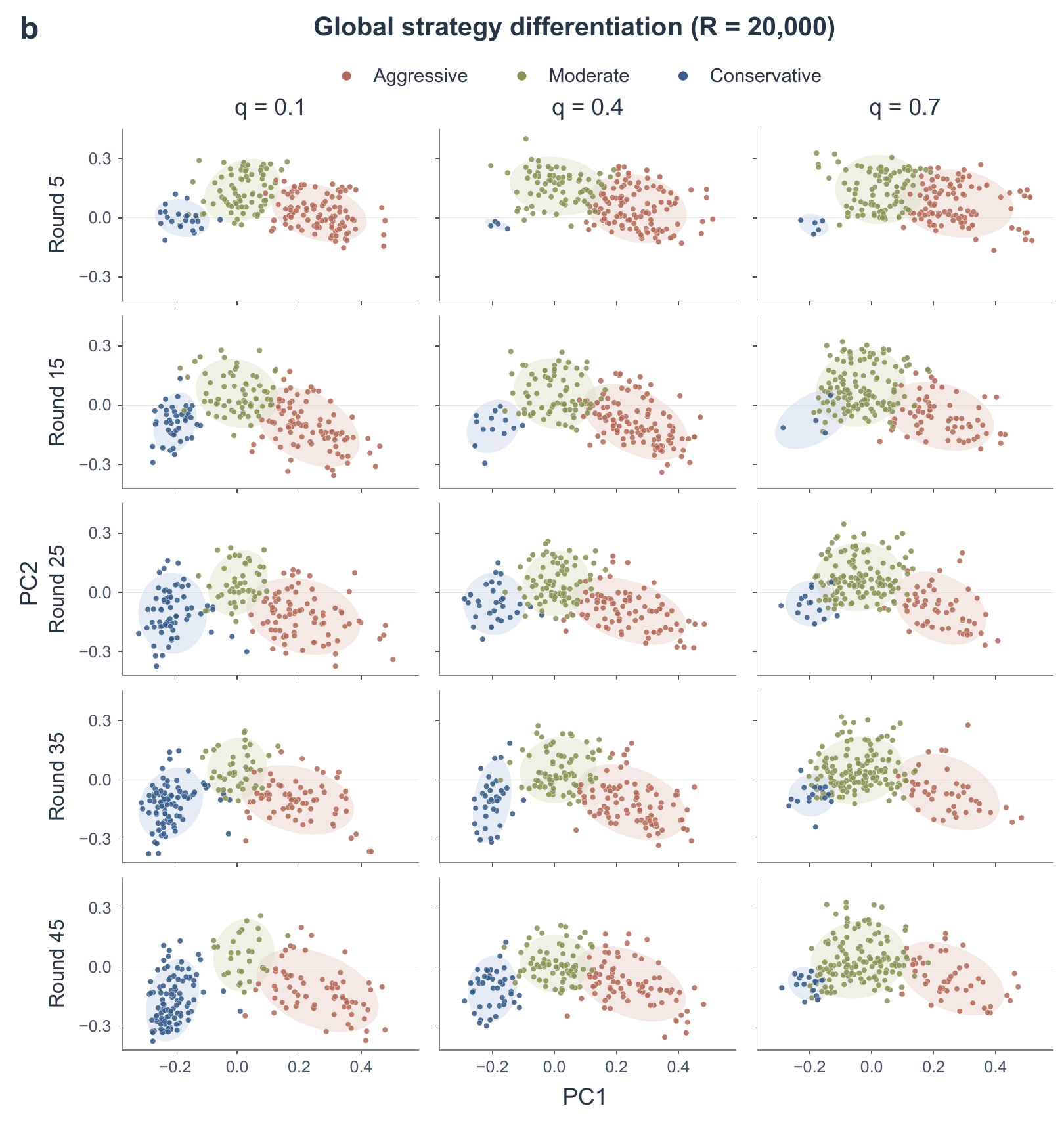}}\par\smallskip
\small (b) Fixed $R=20{,}000$, varying $q$.
\end{minipage}
\caption{Evolution of decision rationales at Rounds 5, 15, 25, 35, and 45. All panels share a global lexical clustering and PCA projection. Colors label aggressive, moderate, and conservative clusters according to their mean investment levels. PCA is a visualization, not an independent test of behavioral separation.}
\Description{Two grids of PCA scatterplots compare strategy evolution across rewards and winner proportions.}
\label{fig:strategy-evolution}
\end{figure*}

\begin{figure*}[!t]
\centering
\begin{minipage}[t]{0.495\textwidth}\centering\vspace{0pt}
\scalebox{1}[0.85]{\includegraphics[width=\linewidth]{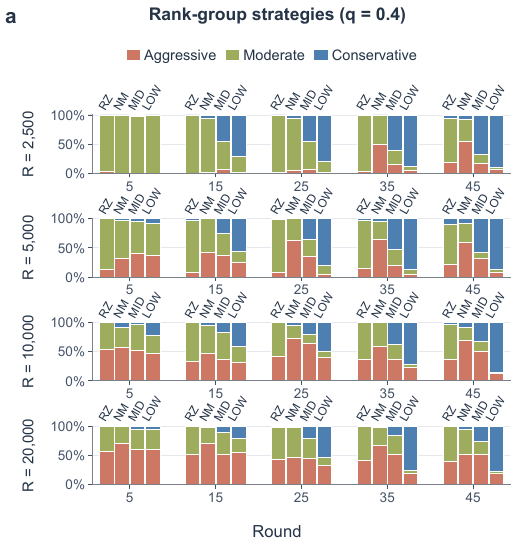}}
\end{minipage}\hfill
\begin{minipage}[t]{0.495\textwidth}\centering\vspace{0pt}
\scalebox{1}[0.85]{\includegraphics[width=\linewidth]{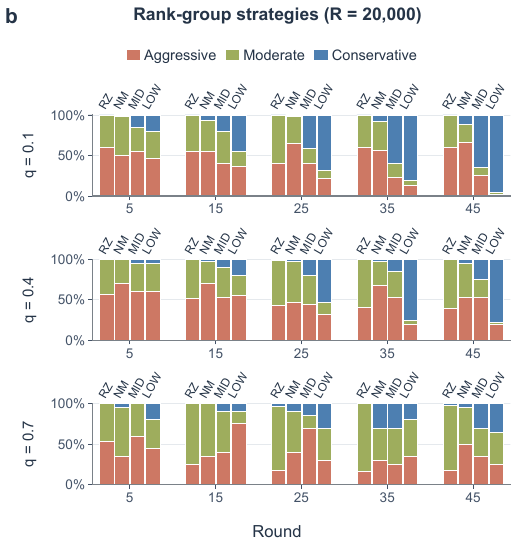}}
\end{minipage}
\caption{Strategy composition across rank bands and rounds. RZ contains the top $qN$ agents; the remaining ranks are divided into NM, MID, and LOW. Bars show within-band proportions, not population counts. Band membership is condition- and round-dependent.}
\Description{Stacked bars compare aggressive, moderate, and conservative strategy proportions across four rank bands.}
\label{fig:rank-strategies}
\end{figure*}

To answer RQ1, we examine aggregate investment relative to the Bayesian adaptive baseline (Figure~\ref{fig:aggregate-investment}), the evolution of aggressive, moderate, and conservative responses (Figure~\ref{fig:strategy-evolution}), and their composition across rank bands (Figure~\ref{fig:rank-strategies}). Together, these analyses connect reward magnitude $R$ and winner proportion $q$ to overcompetition, lying flat, and strategy differentiation.

\noindent\textit{Investment increases with reward magnitude and peaks at intermediate winner coverage within the tested grid.}
At every reward, LLM investment rises from $q=0.1$ to $0.4$ and falls at $0.7$; at each $q$, it increases with $R$. The benchmark also peaks at intermediate coverage but responds more slowly to larger rewards. Thus, the greatest aggregate investment occurs at intermediate coverage within the tested grid, rather than under the narrowest or broadest allocation. Reward magnitude raises the investment level and changes the LLM--benchmark gap.

\noindent\textit{Larger rewards shift populations from lying flat to overcompetition, with the strongest overcompetition at intermediate coverage.}
At $R=2{,}500$, LLM investment lies below the baseline at every tested $q$, indicating lying flat. At $R=5{,}000$, overcompetition appears only at $q=0.4$; the other two configurations remain below the baseline. At $R=10{,}000$ and $20{,}000$, all tested populations overcompete. Gaps increase with $R$ at each $q$, with the strongest overcompetition at $q=0.4$.

\noindent\textit{Larger rewards bring earlier aggressive responses at intermediate coverage, while broader coverage reduces conservative responses at the highest tested reward.}
Figure~\ref{fig:strategy-evolution} depicts coexisting aggressive, moderate, and conservative rationale clusters and their changing composition over time. These descriptive names summarize lexical clusters ordered by associated investment, rather than predefined or independently validated behavioral types (Section~\ref{sec:analysis-measures}).

Reward magnitude primarily affects the timing of differentiation. At fixed $q=0.4$, aggressive responses are already prominent alongside moderate responses at Round~5 under high rewards. Under the lowest reward, early responses are predominantly moderate and aggressive responses emerge later, while conservative responses appear earlier and become more prevalent. Winner coverage, in turn, changes the composition of differentiation. At fixed $R=20{,}000$, narrow coverage produces more conservative responses, especially in later rounds. Broader coverage reduces this conservative component, leaving more agents in aggressive or moderate clusters; the aggressive share itself does not increase monotonically with $q$.

\noindent\textit{Reward design changes where aggressive and conservative responses concentrate across rank bands.}
Figure~\ref{fig:rank-strategies} shows configuration- and round-dependent differences across rank bands. Conservative responses concentrate in LOW/MID; higher-ranked bands generally contain more moderate responses. MID can simultaneously contain substantial aggressive and conservative components. At $q=0.4$, larger rewards increase aggression across bands, often most prominently in NM/MID. At $R=20{,}000$, broader coverage reduces conservatism, especially in lower bands. Late-round aggression is strongest in RZ/NM at $q=0.1$ and more prominent in NM/MID at $q=0.4$; at $q=0.7$, RZ becomes predominantly moderate and LOW more aggressive than under narrow coverage. These are conditional patterns, not a fixed rank ordering; within-band proportions use memberships that vary with $q$ and time.


\subsection{RQ2. The Price of a Chance Is Unevenly Distributed}
\label{sec:individual-costs}
\label{sec:reasoning}

\begin{figure*}[!t]
\centering
\begin{minipage}[t]{0.495\textwidth}\centering\vspace{0pt}
\scalebox{1}[0.85]{\includegraphics[width=\linewidth]{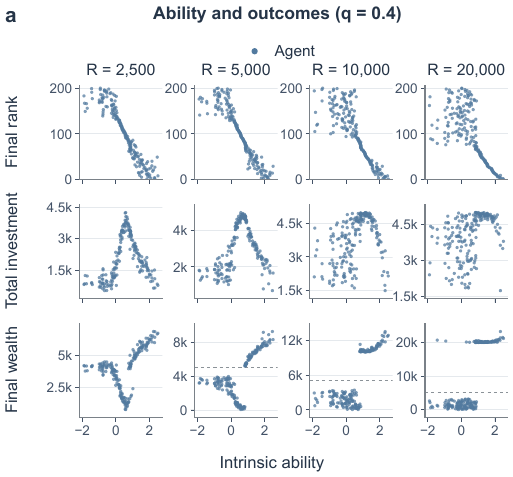}}
\end{minipage}\hfill
\begin{minipage}[t]{0.495\textwidth}\centering\vspace{0pt}
\scalebox{1}[0.85]{\includegraphics[width=\linewidth]{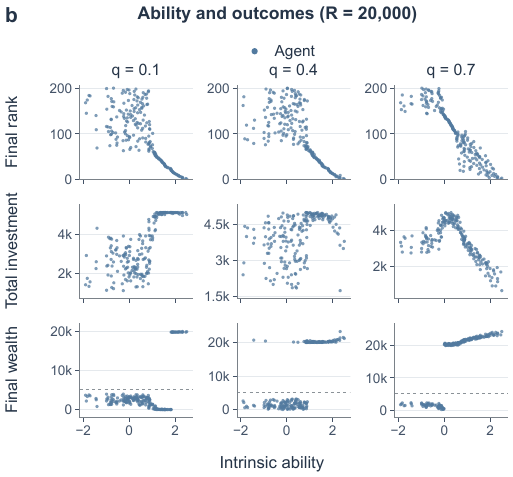}}
\end{minipage}
\caption{Intrinsic ability versus final numerical rank (smaller is better), cumulative investment, and final wealth. Points represent agents. The ability--investment relationship varies with reward magnitude and winner coverage.}
\Description{Scatterplot grids relate intrinsic ability to final rank, investment, and wealth under different rewards and winner proportions.}
\label{fig:ability-outcomes}
\end{figure*}

\noindent\textit{Reward design redistributes the investment burden, leaving some agents close to qualification poorer than other nonwinners.}
Figure~\ref{fig:ability-outcomes} relates intrinsic ability to final rank, investment, and wealth. Ability is the environment's fixed productivity parameter. Higher ability generally predicts better ranks, but not uniformly greater investment. At $q=0.4$, investment peaks at intermediate abilities under lower rewards, while larger rewards broaden the high-investment range. At $R=20{,}000$, high-ability agents invest near their budget under $q=0.1$, whereas broader coverage shifts peak investment toward intermediate abilities; at $q=0.7$, the highest-ability agents maintain favorable ranks with substantially less spending. Across all four rewards at $q=0.4$, some agents just outside the final qualification cutoff invest heavily yet receive no prize, finishing poorer than nonwinners farther from qualification who retain more resources. The costs of pursuing a chance therefore depend on both ability and reward design.

\begingroup
\graphicspath{{sections/}}
\begin{figure*}[!t]
\centering
\scalebox{1}[0.85]{\includegraphics[width=\textwidth]{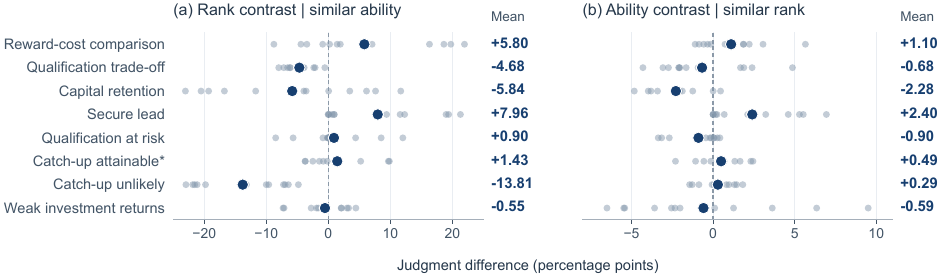}}
\caption{Matched judgment differences in percentage points. (a) Better minus worse rank: ability gap $\leq0.05$ SD and rank gap $\geq20$ places; 29,714 pairs. (b) Higher minus lower ability: rank gap $\leq10$ places and ability gap $\geq1$ SD, with both agents on the same side of the reward boundary; 17,660 pairs. Matching is within the same $(q,R)$ configuration and round, without record reuse within a round. Large dots and numerical annotations show equally weighted means across the 12 reward configurations; small dots show individual configuration differences. Horizontal axis ranges differ between panels. The panels concern different matched populations and do not compare causal importance. The asterisk denotes D2c.}
\Description{Two dot plots show matched judgment differences. Better rank at similar ability is associated with greater security and less unlikely catch-up; ability differences at similar ranks have smaller overall associations that vary with competitive circumstances.}
\label{fig:reasoning43-matching}
\end{figure*}

\noindent\textit{Rank and ability differences accompany distinct judgments under comparable competitive circumstances.}
Figure~\ref{fig:reasoning43-matching} compares judgments associated with rank and ability using two matched populations within the same configuration and decision round. Among 29,714 similar-ability pairs, better-ranked agents express secure leads 7.96 percentage points more often and unlikely catch-up 13.81 points less often; the latter contrast is negative in all 12 configurations. Among 17,660 similar-rank pairs on the same side of the reward boundary, higher-ability agents express secure leads 2.40 points more often, while the overall unlikely-catch-up difference is small (+0.29 points). These averages conceal contextual differences: the secure-lead gap reaches +8.07 points inside the reward zone across nine supported configurations, while weak-return judgments are 6.31 points less frequent outside, with negative differences in 11 of 12 configurations. The ability matches chiefly cover middle- and lower-ability agents (Appendix~\ref{app:reasoning-matching}). These descriptive associations concern different matched populations and do not isolate the causal effects of rank or ability.

\begin{figure}[!tbp]
\centering
\scalebox{1}[0.85]{\includegraphics[width=\linewidth]{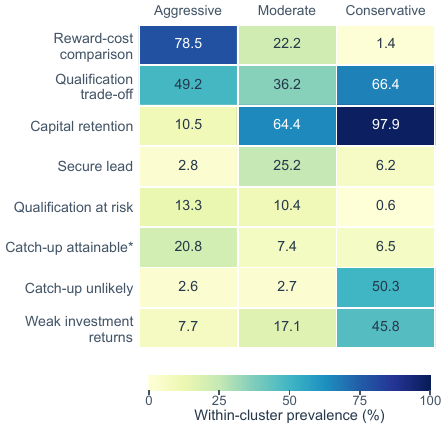}}
\caption{Within-cluster judgment prevalence over 9,600 decisions at four post-feedback snapshots, averaged equally across 12 $(q,R)$ configurations. Labels overlap; * marks attainable catch-up (D2c; Appendix~\ref{app:codebook}).}
\Description{Heatmap comparing eight judgments across aggressive, moderate, and conservative clusters. Aggressive rationales emphasize reward-cost comparisons, conservative rationales emphasize capital retention and unlikely catch-up, and moderate rationales more often express secure leads.}
\label{fig:reasoning43-strategy}
\end{figure}

\noindent\textit{Different strategies accompany distinct judgments about investment returns and qualification prospects.}
Figure~\ref{fig:reasoning43-strategy} compares coded judgments across 9,600 decisions at four post-feedback snapshots, averaged equally across the 12 reward configurations. Broad topic mentions are common across all 107,999 post-feedback rationales (rewards and costs: 99.6\%; competitive position: 98.1\%), yet judgments differ across clusters. Aggressive rationales more often compare rewards with costs (78.5\%, versus 22.2\% moderate and 1.4\% conservative). They also express threatened qualification or attainable catch-up more often than conservative rationales. Conservative rationales emphasize capital retention (97.9\%), unlikely catch-up (50.3\%), and weak investment returns (45.8\%). Secure-lead judgments are most common in moderate rationales (25.2\%, versus 2.8\% aggressive and 6.2\% conservative). Qualification-probability trade-offs occur in all three clusters, and the principal contrasts persist within the same configuration and round. Because clustering and coding use the same rationales, these patterns characterize expressed reasoning rather than independently establish a causal mechanism.

\endgroup

\subsection{RQ3. Greater Investment Need Not Improve Selection}
\label{sec:collective-outcomes}
\begin{figure}[!tbp]
\centering
\scalebox{1}[0.85]{\includegraphics[width=\linewidth]{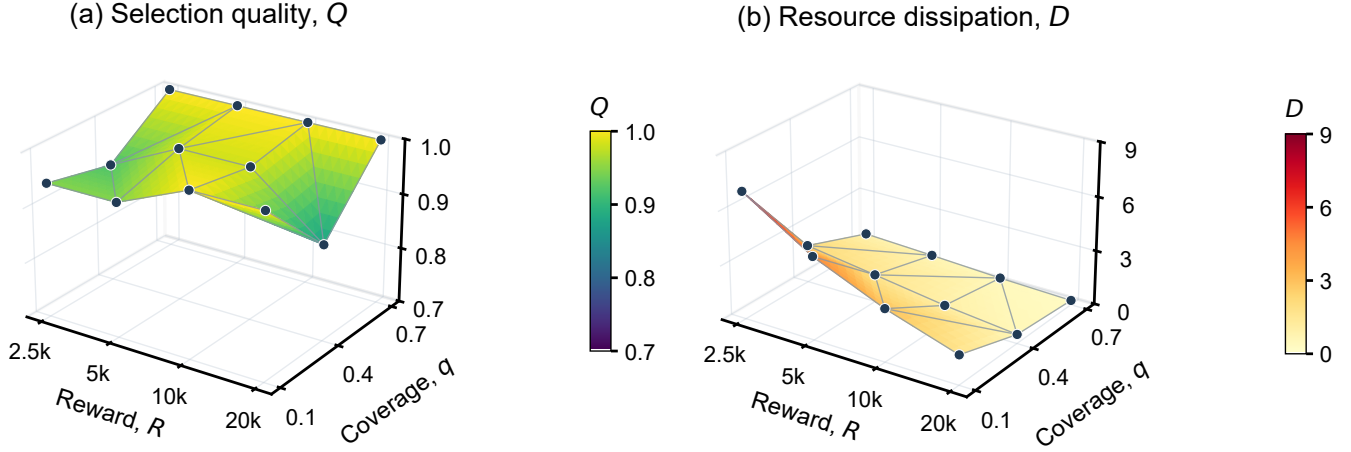}}
\caption{LLM selection quality $Q$ (higher is better) and prize-normalized dissipation $D$ (lower is better): 12 main $(q,R)$ configurations, $\sigma_c=1$. Dark dots are observations; surfaces interpolate piecewise-linearly in $(\log_2 R,q)$, with main-grid triangulation shown. Axis/color ranges: $Q\in[0.7,1]$, $D\in[0,9]$. Surfaces add no observations; across varying budgets, lower $D$ need not mean lower absolute investment.}
\Description{Separate three-dimensional views show selection quality and normalized resource dissipation over winner coverage and reward magnitude, with the 12 main reward-grid configurations marked. Selection quality is not uniformly increasing in reward; normalized dissipation falls across the main grid as coverage or reward increases.}
\label{fig:qd-landscape}
\end{figure}

\begin{figure}[!tbp]
\centering
\scalebox{1}[0.85]{\includegraphics[width=0.9\linewidth]{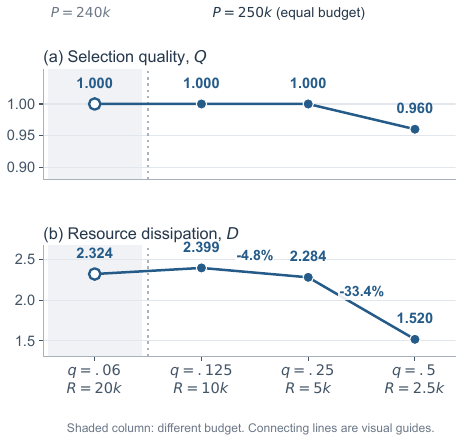}}
\caption{LLM selection quality and normalized dissipation across four allocations. Shading marks the $P=240{,}000$ reference; the other three share $P=250{,}000$. Lines are visual guides. Percentage reductions concern consecutive equal-budget allocations; k denotes thousands.}
\Description{LLM-only outcomes show that equal-budget reallocation first reduces investment without lowering selection quality, while further savings accompany a small quality reduction. The different-budget reference is shaded.}
\label{fig:fixed-budget-outcomes}
\end{figure}

\begin{figure}[!tbp]
\centering
\scalebox{1}[0.85]{\includegraphics[width=0.9\linewidth]{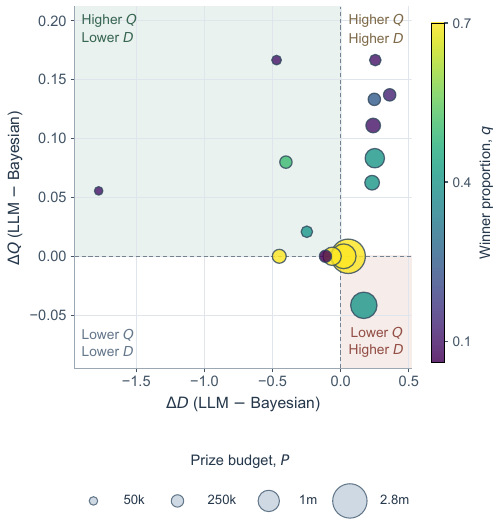}}
\caption{LLM minus Bayesian outcomes across all 16 configurations: $\Delta Q=Q_\mathrm{LLM}-Q_\mathrm{Bayesian}$ and $\Delta D=D_\mathrm{LLM}-D_\mathrm{Bayesian}$. LLM outperforms the baseline in quadrant II (upper-left: higher $Q$, lower $D$) and underperforms in quadrant IV (lower-right: lower $Q$, higher $D$). Color encodes $q$; marker area increases linearly with prize budget $P$, with a small visibility offset. Points on $\Delta Q=0$ can overlap.}
\Description{A single-column four-quadrant bubble plot compares selection quality and resource dissipation for all 16 configurations. Colors encode winner proportions and marker sizes encode prize budgets.}
\label{fig:qd-deltas}
\end{figure}

The unequal costs of participation raise a collective question: what does the price of a chance achieve for the collective (RQ3)? We assess selection quality $Q$ and prize-normalized dissipation $D$ (Section~\ref{sec:analysis-measures}) across the reward landscape (Figure~\ref{fig:qd-landscape}), fixed-budget allocations (Figure~\ref{fig:fixed-budget-outcomes}), and Bayesian-relative outcomes (Figure~\ref{fig:qd-deltas}).

\noindent\textit{Greater investment does not consistently improve ability selection.}
Figure~\ref{fig:qd-landscape} reveals distinct patterns across the 12 main reward configurations. Selection quality is highest under broad winner coverage and also reaches its maximum under narrow coverage with sufficiently large rewards. At intermediate coverage, however, increasing rewards does not consistently improve selection: quality falls at the largest reward despite stronger aggregate investment. Normalized dissipation, in contrast, decreases as either coverage or reward increases. Broad coverage and large rewards therefore combine high selection quality with low dissipation per unit of prize budget within the tested grid. Because the budget itself varies, this does not establish lower absolute investment; we next hold the budget fixed.

\noindent\textit{Equal-budget reallocation initially reduces investment without lowering selection quality, but further savings accompany a quality loss.}
Among the three $P=250{,}000$ allocations, broader coverage with smaller prizes reduces dissipation (Figure~\ref{fig:fixed-budget-outcomes}). The $q=0.25$ allocation preserves maximal selection quality with less investment than narrower coverage; expanding to $q=0.5$ yields larger savings with a modest quality loss. Because changing $q$ changes $K$, these quality comparisons concern each allocation's target set rather than recovery of a single fixed group. Allocation choice therefore depends on the desired selection coverage and conservation priorities. The shaded $P=240{,}000$ reference is excluded from strict equal-budget savings comparisons.

\noindent\textit{The selection and investment advantages of LLM populations over the baseline depend on reward design.}
Lower rewards tend to place LLM outcomes in quadrant II (higher $Q$, lower $D$; Figure~\ref{fig:qd-deltas}). With larger rewards, narrow/intermediate coverage more often shifts to quadrant I (higher quality and cost), while intermediate coverage at the largest reward falls in quadrant IV (lower quality, higher cost). Broad coverage retains equal selection quality, but its cost advantage at low rewards reverses at high rewards. Relative performance therefore depends on reward design.

\section{Discussion}
\label{sec:discussion}
\paragraph{Collective behavior emerges through differentiated adaptation.}
The aggregate investment profile reflects the coexistence of escalation and resource preservation rather than a uniform population response. Their changing composition connects individual adaptation to emergent collective behavior~\cite{macy2002actors}. Stated reasoning offers an interpretive bridge: agents discuss similar considerations but express different judgments about qualification security, catch-up prospects, and investment returns. Rank-associated differences remain among agents with similar abilities, while ability-associated differences at comparable ranks depend on competitive circumstances. Together, these findings suggest that reward structures shape collective behavior through differentiated assessments of opportunities and costs. They do not, however, establish stated explanations as causal mechanisms. Understanding the aggregate pattern therefore requires examining both how much agents invest and how their responses differ across competitive positions.

\paragraph{Competition design should align incentives with collective objectives.}
Mobilizing greater investment does not necessarily improve selection. Mechanism evaluation should therefore consider selection quality and resource use alongside aggregate effort. Fixed-budget comparisons reveal opportunities to preserve selection quality while reducing expenditure, as well as trade-offs when pursuing further savings. Reward allocations should consequently reflect the designer's priorities rather than maximize competitive intensity alone. Two distinctions matter when interpreting these choices. First, lower prize-normalized dissipation does not imply lower absolute expenditure when prize budgets differ; $D$ should be read alongside total investment $I$ and budget $P$. Second, changing winner coverage changes the target top-$K$ ability set, so $Q$ evaluates selection relative to each allocation's target. The framework thus supports comparing the consequences of alternative designs, without prescribing a universally preferred allocation.

\paragraph{Scope and interpretation.}
These conclusions concern the tested environment and parameter grid. In particular, $q=0.4$ is an observed investment peak among the tested coverage levels, not a continuous optimum or an optimal competition mechanism. Other rules and parameter settings may alter strategy composition and outcome trade-offs. Likewise, benchmark-relative gaps describe differences between the implemented decision systems under their information assumptions, not departures from a universal rational optimum. Finally, $Q$ and $D$ capture ability selection and resource use rather than comprehensive social welfare. Applying these insights to specific Web-mediated communities requires adapting the simulated rules to their ranking and reward mechanisms and validating behavioral patterns against empirical evidence~\cite{gao2024survey,larooij2026validation}.

\section{Conclusion}
\label{sec:conclusion}
This study connects reward design, individual adaptation, and collective outcomes in repeated competition among LLM agents. Reward magnitude raises investment, while winner coverage produces an inverted-U aggregate response within the tested grid. These patterns emerge alongside differentiated strategies across ranks and abilities, with stated judgments linking engagement to perceived qualification prospects and investment returns. Greater expenditure does not consistently improve ability selection; equal-budget comparisons instead reveal opportunities to reduce investment while preserving selection quality, as well as quality--cost trade-offs. Evaluating competition mechanisms therefore requires considering reward allocation, heterogeneous responses, and resource use together. The framework supports controlled investigation of these relationships, with replication and empirical validation needed to establish their robustness and relevance beyond the simulated setting.
\FloatBarrier
\bibliographystyle{plainnat}
\bibliography{references-arxiv}

@article{lazear1981tournaments,
  author = {Lazear, Edward P. and Rosen, Sherwin},
  title = {Rank-Order Tournaments as Optimum Labor Contracts},
  journal = {Journal of Political Economy},
  year = {1981}, volume = {89}, number = {5}, pages = {841--864},
  doi = {10.1086/261010}
}

@article{dechenaux2015survey,
  author = {Dechenaux, Emmanuel and Kovenock, Dan and Sheremeta, Roman M.},
  title = {A survey of experimental research on contests, all-pay auctions and tournaments},
  journal = {Experimental Economics},
  year = {2015}, volume = {18}, number = {4}, pages = {609--669},
  doi = {10.1007/s10683-014-9421-0}
}

@article{macy2002actors,
  author = {Macy, Michael W. and Willer, Robert},
  title = {From Factors to Actors: Computational Sociology and Agent-Based Modeling},
  journal = {Annual Review of Sociology},
  year = {2002}, volume = {28}, pages = {143--166},
  doi = {10.1146/annurev.soc.28.110601.141117}
}

@article{mei2024turing,
  author = {Mei, Qiaozhu and Xie, Yutong and Yuan, Walter and Jackson, Matthew O.},
  title = {A {Turing} test of whether {AI} chatbots are behaviorally similar to humans},
  journal = {Proceedings of the National Academy of Sciences},
  year = {2024}, volume = {121}, number = {9}, pages = {e2313925121},
  doi = {10.1073/pnas.2313925121}
}

@article{akata2025games,
  author = {Akata, Elif and Schulz, Lion and Coda-Forno, Julian and Oh, Seong Joon and Bethge, Matthias and Schulz, Eric},
  title = {Playing repeated games with large language models},
  journal = {Nature Human Behaviour},
  year = {2025}, volume = {9}, pages = {1380--1390},
  doi = {10.1038/s41562-025-02172-y}
}

@article{gao2024survey,
  author = {Gao, Chen and Lan, Xiaochong and Li, Nian and Yuan, Yuan and Ding, Jingtao and Zhou, Zhilun and Xu, Fengli and Li, Yong},
  title = {Large language models empowered agent-based modeling and simulation: a survey and perspectives},
  journal = {Humanities and Social Sciences Communications},
  year = {2024}, volume = {11}, pages = {1259},
  doi = {10.1057/s41599-024-03611-3}
}

@article{larooij2026validation,
  author = {Larooij, Maik and T{\"o}rnberg, Petter},
  title = {Validation is the central challenge for generative social simulation: a critical review of {LLMs} in agent-based modeling},
  journal = {Artificial Intelligence Review},
  year = {2026}, volume = {59}, pages = {15},
  doi = {10.1007/s10462-025-11412-6}
}

@article{boosey2020information,
  author  = {Boosey, Luke and Brookins, Philip and Ryvkin, Dmitry},
  title   = {Information disclosure in contests with endogenous entry: An experiment},
  journal = {Management Science},
  volume  = {66},
  number  = {11},
  pages   = {5128--5150},
  year    = {2020},
  doi = {10.1287/mnsc.2019.3488}
}

@article{cason2020winner,
  author  = {Cason, Timothy N. and Masters, William A. and Sheremeta, Roman M.},
  title   = {Winner-take-all and proportional-prize contests: Theory and experimental results},
  journal = {Journal of Economic Behavior \& Organization},
  volume  = {175},
  pages   = {314--327},
  year    = {2020}
}

@article{chaocheng2023involution,
  author  = {He, Chaocheng and Huang, Qian and Li, Xinru and Zuo, Renxian and Liu, Fuzhen and Wei, Yuchi and Wu, Jiang},
  title   = {Involution-cooperation-lying flat game on a network-structured population in the group competition},
  journal = {IEEE Transactions on Computational Social Systems},
  volume  = {11},
  number  = {2},
  pages   = {2160--2178},
  year    = {2024},
  doi = {10.1109/TCSS.2023.3287772}
}

@inproceedings{park2023generative,
  author    = {Park, Joon Sung and O'Brien, Joseph and Cai, Carrie Jun and Morris, Meredith Ringel and Liang, Percy and Bernstein, Michael S.},
  title     = {Generative agents: Interactive simulacra of human behavior},
  booktitle = {Proceedings of the 36th Annual ACM Symposium on User Interface Software and Technology},
  numpages  = {22},
  year      = {2023},
  doi = {10.1145/3586183.3606763},
  publisher = {Association for Computing Machinery},
  address = {New York, NY},
  articleno = {2}
}

@article{wang2021replicator,
  author  = {Wang, Chaoqian and Hui, Kangshuo},
  title   = {Replicator dynamics for involution in an infinite well-mixed population},
  journal = {Physics Letters A},
  volume  = {420},
  pages   = {127759},
  year    = {2021},
  doi = {10.1016/j.physleta.2021.127759}
}

@article{wang2022modeling,
  author  = {Wang, Chaoqian and Huang, Chaochao and Pan, Qiuhui and He, Mingfeng},
  title   = {Modeling the social dilemma of involution on a square lattice},
  journal = {Chaos, Solitons \& Fractals},
  volume  = {158},
  pages   = {112092},
  year    = {2022},
  doi = {10.1016/j.chaos.2022.112092}
}

@inproceedings{wu2024autogen,
  author = {Wu, Qingyun and Bansal, Gagan and Zhang, Jieyu and Wu, Yiran and Li, Beibin and Zhu, Erkang and Jiang, Li and Zhang, Xiaoyun and Zhang, Shaokun and Liu, Jiale and Awadallah, Ahmed and White, Ryen W. and Burger, Doug and Wang, Chi},
  title = {{AutoGen}: Enabling Next-Gen {LLM} Applications via Multi-Agent Conversations},
  booktitle = {First Conference on Language Modeling},
  year = {2024},
  url = {https://www.microsoft.com/en-us/research/wp-content/uploads/2023/08/LLM_agent.pdf}
}

@inproceedings{hong2024metagpt,
  author = {Hong, Sirui and Zhuge, Mingchen and Chen, Jonathan and Zheng, Xiawu and Cheng, Yuheng and Wang, Jinlin and Zhang, Ceyao and Wang, Zili and Yau, Steven and Lin, Zijuan and Zhou, Liyang and Ran, Chenyu and Xiao, Lingfeng and Wu, Chenglin and Schmidhuber, J{\"u}rgen},
  title = {{MetaGPT}: Meta Programming for a Multi-Agent Collaborative Framework},
  booktitle = {The Twelfth International Conference on Learning Representations},
  year = {2024},
  url = {https://proceedings.iclr.cc/paper_files/paper/2024/hash/6507b115562bb0a305f1958ccc87355a-Abstract-Conference.html}
}

@inproceedings{zhu2025automated,
  author = {Zhu, Shenzhe and Sun, Jiao and Nian, Yi and South, Tobin and Pentland, Alex and Pei, Jiaxin},
  title = {The Automated but Risky Game: Modeling Agent-to-Agent Negotiations and Transactions in Consumer Markets},
  booktitle = {Proceedings of the Natural Legal Language Processing Workshop 2025},
  year = {2025},
  pages = {16},
  publisher = {Association for Computational Linguistics},
  doi = {10.18653/v1/2025.nllp-1.2},
  url = {https://aclanthology.org/2025.nllp-1.2/}
}

@article{ashery2025conventions,
  author = {Ashery, Ariel Flint and Aiello, Luca Maria and Baronchelli, Andrea},
  title = {Emergent social conventions and collective bias in {LLM} populations},
  journal = {Science Advances},
  year = {2025},
  volume = {11},
  number = {20},
  pages = {eadu9368},
  doi = {10.1126/sciadv.adu9368}
}

@article{piao2026polarization,
  author = {Piao, Jinghua and Lu, Zhihong and Gao, Chen and Xu, Fengli and Hu, Qinghua and Santos, Fernando P. and Li, Yong and Evans, James},
  title = {Emergence of polarization in networks of large language model agents},
  journal = {Nature Communications},
  year = {2026},
  doi = {10.1038/s41467-026-78228-y}
}
\appendix
\section{Additional Method Details}
\label{app:methods}
\subsection{Bayesian Qualification Approximation and Decision Rule}
\label{app:benchmark}
For each candidate integer investment $x\in\{0,\ldots,100\}$, an agent approximates its probability of qualifying using
\begin{equation}
 \widehat p_{it}(x)=\Phi\!\left(\frac{\beta(x-k)h_t-(r_i-K)g}{\widehat\sigma_i\sqrt{H_t}+10^{-8}}\right),
 \label{eq:benchmark-probability}
\end{equation}
where $H_t=T-t+1$, $h_t=(1-\gamma^{H_t})/(1-\gamma)$, and $\Phi$ is the standard normal cumulative distribution function. Here, $r_i$ is the latest available rank, $\beta$ is estimated investment productivity, $k$ is the estimated cutoff investment, and $g$ converts rank distance into a cumulative score deficit. The implemented discount factor is $\gamma=0.7$. Averaging qualification probabilities over 30 sampled parameter settings, the agent chooses
\begin{equation}
 x^{\mathrm{B}}_{it}\in\arg\max_{x\in\{0,\ldots,100\}}\left\{-x+R\,\mathbb E[\widehat p_{it}(x)]\right\}.
 \label{eq:benchmark-decision}
\end{equation}
Productivity and noise estimates use five recent observations; rank-dependent estimates update with published rankings. Guaranteed rewards yield zero investment.

This approximate current-cost/terminal-reward objective does not solve the dynamic game or compute an exact population posterior. Information initialization differs from the LLM agents (Section~\ref{sec:simulation-framework}).

\subsection{Rationale Clustering and Visualization}
\label{app:clustering}
We pool rationales at Rounds 5, 15, 25, 35, and 45 across the 12 main configurations. Pattern normalization replaces selected reward, rank, investment, score, and timing expressions with placeholders. Shared TF--IDF features use English stop-word removal, 1--3-grams, at most 600 features, and document frequencies from 3 to 80\%. K-means uses three clusters (seed 42; 10 initializations), named by descending associated mean investment. PCA provides a two-dimensional visualization; clustering uses the original features. Labels are descriptive clusters, not investment thresholds or validated psychological types.

Strategy snapshots use ranks saved with that round; reasoning comparisons use pre-decision published ranks. Rank bands vary with time and $q$. At $q=1$, ranking does not affect eligibility and $Q$ is undefined.

\subsection{LLM Decision Prompt and Feedback Sequence}
\label{app:llm-prompt}
Figure~\ref{fig:llm-prompt-pseudocode} summarizes the implemented template; braces denote supplied fields. True ability is not numerically supplied, and no separate periodic reflection call runs.

\begin{figure}[!tbp]
\centering
\begin{minipage}{\linewidth}
\small
\hrule
\vspace{7pt}
\begin{verbatim}
BUILD_PROMPT(agent i, round t)
ROLE: Investor {i}; {N} players.
GOAL: Maximize retained capital + terminal reward.
RULES:
  Receive 100 units/round; invest an integer 0-100.
  Retain the unspent part of each round's endowment for final wealth; it cannot be reinvested.
  Scores use relative investment, ability, noise; cumulative average scores determine ranks.
  Publish ranks every 5 rounds. Top {q*100}% earn {R} at the end; others earn zero.
  Scores do not directly add to wealth.
INPUTS:
  Horizon, round, supplied remaining-round count; retained capital and cumulative investment;
  own investments/scores from up to 5 recent rounds;
  latest published rank and cumulative average score, or "not available" before publication.
GUIDANCE:
  Identify information relevant to final wealth;
  review recent investments and resulting scores.
  Assess qualification prospects and whether further investment can improve them.
  Compare the required expenditure with the potential terminal reward and retained capital.
  Infer scoring ability from observed results when evaluating additional
  investment productivity.
  Choose investment; explain in 1-2 sentences.
RETURN JSON:
  {"thinking_process": "...", "investment_amount": integer 0-100,
   "reason": "1-2 sentence explanation"}
\end{verbatim}
\vspace{4pt}
\hrule
\end{minipage}
\caption{Condensed decision prompt. Reasoning analysis uses the saved \texttt{reason} field.}
\Description{Prompt pseudocode states rules, available information, decision guidance, and JSON output.}
\label{fig:llm-prompt-pseudocode}
\end{figure}

All decisions precede scoring. Own scores, rolling histories, and retained capital then update. Ranks and cumulative average scores refresh every five rounds; intervening decisions retain the previous published signals. Round~6 is the first decision after ranking feedback.

\section{Supplementary Aggregate Investment Results}
\label{app:legacy-models}
\subsection{Model Variation in Archived Experiments}
\begin{figure}[!tbp]
\centering
\includegraphics[width=0.88\linewidth]{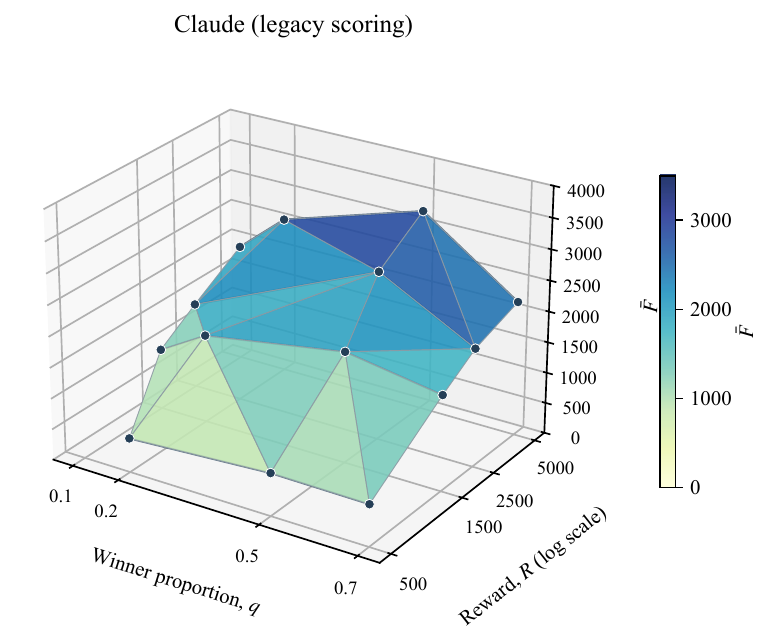}
\caption{Archived Claude investment: 14 observed configurations, with piecewise-linear interpolation in $(q,\log_{10}R)$. Earlier scoring includes cost subtraction.}
\Description{Claude investment peaks at intermediate winner coverage in the archived scoring environment.}
\label{fig:legacy-models}
\end{figure}
Figure~\ref{fig:legacy-models} reports 14 archived Claude configurations ($N=200$, $T=50$). Investment peaks at the tested $q=0.5$ for each reward and declines at $q=0.7$. These runs use an earlier scoring rule with cost subtraction, so they illustrate a related environment rather than a controlled model replacement. Exact model versions are unavailable; no benchmark-relative or collective-outcome comparisons use these runs.

\subsection{Ability Dispersion}
\label{app:ability-dispersion}
\begin{figure}[!tbp]
\centering
\includegraphics[width=0.86\linewidth]{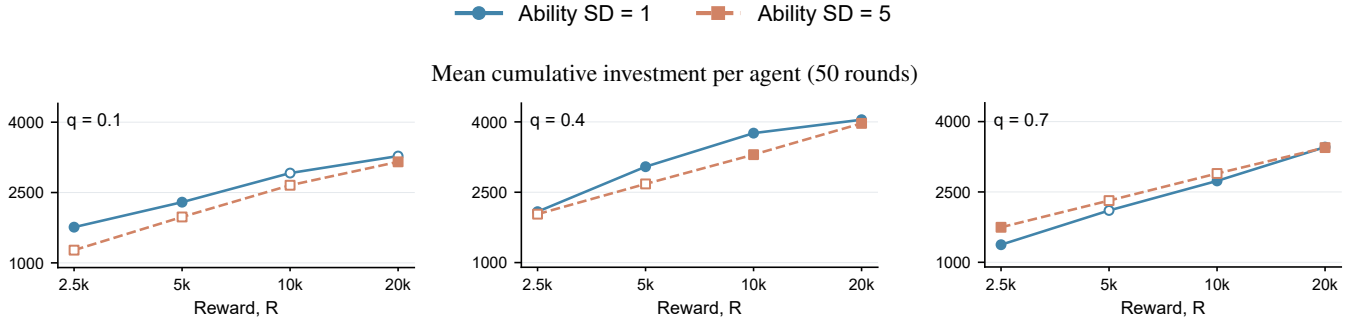}
\caption{Ability-dispersion comparisons at fixed $q$. Each point is mean cumulative investment per agent over 50 rounds. Lines connect observed rewards; hollow markers flag record inconsistencies.}
\Description{Three horizontally arranged panels fix winner coverage at 0.1, 0.4, and 0.7 and compare investment under ability standard deviations 1 and 5 as rewards increase.}
\label{fig:ability-dispersion}
\end{figure}

Figure~\ref{fig:ability-dispersion} compares ability dispersion across the main reward grid. Investment is higher under $\sigma_c=1$ at $q=0.1$ and $0.4$, whereas $\sigma_c=5$ gives higher investment at $q=0.7$ except at the largest reward, where the curves nearly coincide. Thus, the level difference depends on coverage.

Totals are reconstructed from 50 round-level decisions per agent, with one run per setting. Ten of 24 runs have inconsistent saved cumulative totals or feedback fields (three under $\sigma_c=1$, seven under $\sigma_c=5$; hollow markers). Reconstruction corrects the plotted totals but not the feedback experienced by agents; this comparison is descriptive.

\section{Reasoning Coding and Supplementary Label Results}
\label{app:reasoning}

\begin{figure*}[!t]
\centering
\includegraphics[width=\textwidth]{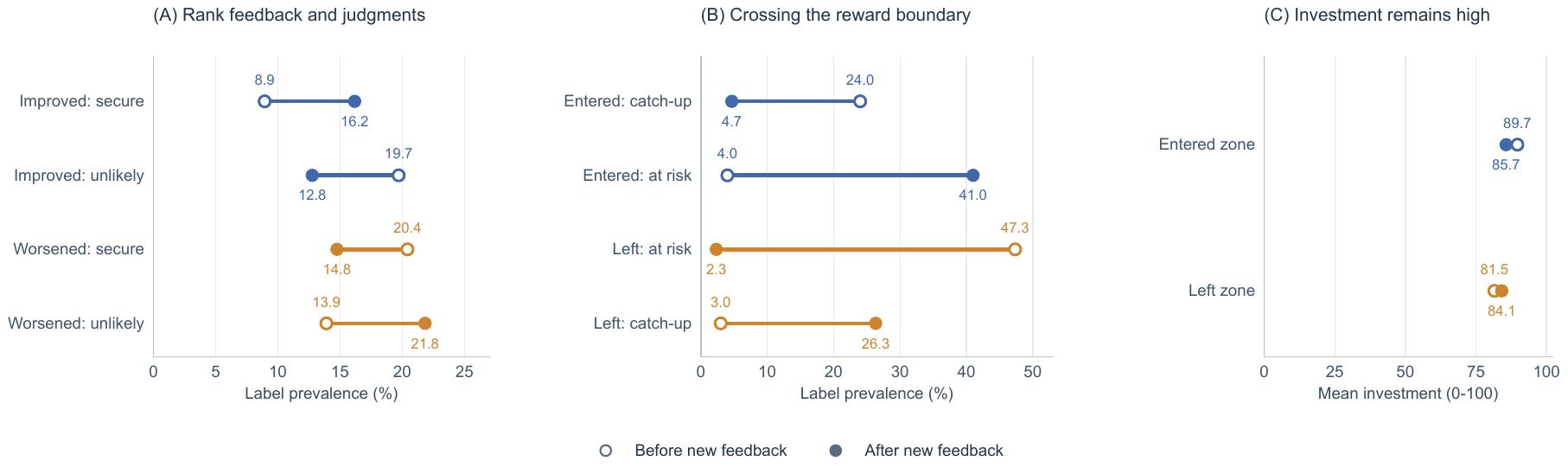}
\caption{Within-agent comparisons around bundled ranking and score feedback. Hollow/filled markers show decisions at $k$/$k+1$, $k=10,15,\ldots,45$. (A) Rank improvements and deteriorations of at least ten places (3,110 and 3,532 events). (B-C) Entry into and exit from the reward zone (300 each). Repeated events are descriptive, not independent experimental replications.}
\Description{Within-agent changes in coded judgments and investment around published feedback.}
\label{fig:reasoning-feedback}
\end{figure*}
\subsection{Coding and Aggregation Protocol}
The rule-based codebook comprises five dimensions and 25 overlapping substantive labels (Table~\ref{tab:reasoning-codebook}). One unusable output is excluded, leaving 119,999 rationales, including 107,999 post-feedback decisions (Rounds 6--50). Group rates are averaged equally across configurations. Codes reflect explicit expressions, not internal attention; unmatched categories are excluded. D1/D3/D5 permit multiple labels; conflicting D2 matches remain flagged. D4 economizing/withdrawal require secure-lead/unlikely-catch-up evidence, respectively, so their co-occurrence with D2 is not independent evidence. Zero investment alone does not establish withdrawal; D5 concerns stated beliefs about others.

\subsection{Codebook and Coverage}
\label{app:codebook}

\begin{table*}[!t]
\centering\footnotesize
\setlength{\tabcolsep}{3pt}
\renewcommand{\arraystretch}{1.0}
\caption{Reasoning codebook and coverage: five dimensions, 25 overlapping labels, and counts/rates over 119,999 rationales. Rank/ability spreads are maximum minus minimum marginal post-feedback group rates (percentage points), averaged equally across configurations; they do not adjust for rank--ability associations or test significance. Sparse matches do not establish absence of consideration.}
\label{tab:reasoning-codebook}\label{tab:reasoning-coverage}
\begin{tabular}{@{}lp{0.19\textwidth}p{0.42\textwidth}rrrr@{}}
\hline
Code & Label & Operational meaning & $n$ & Rate (\%) & Rank & Ability \\
\hline
\multicolumn{7}{@{}l}{\textbf{D1}: Reward--cost reasoning} \\
D1a & Reward--cost comparison & Compares reward with expenditure, savings, or capital at risk. & 34,885 & 29.07 & 30.87 & 8.14 \\
D1b & Qualification-probability trade-off & Weighs investment costs/gains against qualification probability. & 53,246 & 44.37 & 40.70 & 30.51 \\
D1c & Capital retention & Emphasizes retaining resources or avoiding unnecessary spending. & 72,924 & 60.77 & 39.58 & 13.39 \\
D1d & Exploration value & Values investment for testing or acquiring information. & 2,376 & 1.98 & 0.76 & 0.54 \\
\multicolumn{7}{@{}l}{\textbf{D2}: Perceived competitive position} \\
D2a & Secure lead & Judges position safe or qualification buffer sufficient. & 18,770 & 15.64 & 31.32 & 41.54 \\
D2b & Qualification at risk & Judges eligibility vulnerable or its buffer narrow. & 9,495 & 7.91 & 25.49 & 9.59 \\
D2c & Attainable catch-up & Judges a deficit recoverable or qualification attainable. & 10,962 & 9.14 & 24.77 & 3.26 \\
D2d & Unlikely catch-up & Judges qualification or recovery improbable. & 17,093 & 14.24 & 43.15 & 29.58 \\
D2e & Explicit uncertainty & Expresses uncertainty about position or qualification. & 9,458 & 7.88 & 2.59 & 1.45 \\
\multicolumn{7}{@{}l}{\textbf{D3}: Investment efficacy and ability} \\
D3a & Effective investment & Attributes performance or qualification gains to investment. & 5,214 & 4.35 & 4.23 & 2.83 \\
D3b & Weak investment returns & Judges further spending ineffective or of limited benefit. & 26,363 & 21.97 & 19.95 & 14.89 \\
D3c & Uncertain effectiveness & Expresses uncertainty about investment effects. & 7,389 & 6.16 & 4.35 & 3.77 \\
D3d & Ability advantage & Attributes an advantage explicitly to own ability. & 58 & 0.05 & 0.02 & 0.00 \\
D3e & Ability disadvantage & Attributes a disadvantage explicitly to own ability. & 29 & 0.02 & 0.07 & 0.06 \\
D3f & Noise attribution & Attributes outcomes to noise or randomness. & 2,586 & 2.16 & 2.33 & 2.06 \\
\multicolumn{7}{@{}l}{\textbf{D4}: Action intent} \\
D4a & Catch-up intent & Proposes investing to improve rank or reach eligibility. & 13,755 & 11.46 & 24.88 & 17.23 \\
D4b & Defend eligibility & Proposes spending to protect eligibility. & 26,954 & 22.46 & 50.17 & 35.64 \\
D4c & Economizing & Proposes saving resources given a perceived secure lead. & 14,320 & 11.93 & 22.94 & 32.44 \\
D4d & Withdrawal & Proposes disengaging given perceived inability to catch up. & 14,525 & 12.10 & 37.08 & 25.23 \\
D4e & Testing & Proposes a trial or exploratory action. & 960 & 0.80 & 0.43 & 0.25 \\
D4f & Maintain strategy & Proposes continuing the current strategy. & 6,481 & 5.40 & 5.85 & 4.50 \\
\multicolumn{7}{@{}l}{\textbf{D5}: Others' behavior} \\
D5a & Follow others' high effort & Refers to matching others' high investment. & 7 & 0.01 & 0.00 & 0.00 \\
D5b & Preempt others' escalation & Anticipates and responds to others' increased effort. & 10 & 0.01 & 0.02 & 0.02 \\
D5c & Use others' low effort & Refers to exploiting others' low investment. & 0 & 0.00 & 0.00 & 0.00 \\
D5d & Relative-effort assessment & Compares own effort with others' effort. & 114 & 0.10 & 0.41 & 0.28 \\
\hline
\end{tabular}
\end{table*}

\paragraph{Additional labels.}
\label{app:additional-labels}
Rounds 1--5 contain 66.08\% of exploration and 75.62\% of testing matches. Explicit uncertainty has smaller marginal group gaps than capital retention and weak returns. Ability self-assessments and D5 have sparse explicit matches, not proven absence of consideration. Noise attribution covers 2.16\% of decisions and accompanies mean investment 77.68. D2c/D3a retain semantic-boundary cases.

\subsection{Strategy Alignment and Matching Protocol}
\label{app:reasoning-matching}
All 12,000 snapshot investments agree across clustering and coding files. Strategy comparisons use 9,600 snapshots from Rounds 15--45. Within-condition/round comparisons preserve the main directions across all 12 conditions: aggressive rationales show more reward--cost comparisons and less capital retention/unlikely catch-up than conservative rationales; moderate rationales show more secure leads than either cluster.

Both matching designs pair post-feedback records within the same condition and decision round. Ability is standardized using the 200 fixed abilities in each condition. No record is reused within a condition--round matching, although agents recur across rounds. For rank comparisons, candidates have ability gaps $\leq0.05$ SD and rank gaps $\geq20$ places. Greedy matching prioritizes the smallest ability gap, then the largest rank gap. The 29,714 pairs involve 2,044 condition-specific agent trajectories, with mean gaps of 0.0137 SD and 45.0 places. Contrasts are better minus worse rank. Among pairs both inside the reward zone, the secure-lead contrast is +22.58 points across eight supported conditions; among pairs both outside, the unlikely-catch-up contrast is $-21.77$ across 12.

For ability comparisons, candidates have rank gaps $\leq10$ places, ability gaps $\geq1$ SD, and positions on the same side of the reward boundary. Greedy matching prioritizes the smallest rank gap, then the largest ability gap. The 17,660 pairs involve 1,750 trajectories, with mean gaps of 2.27 places and 1.40 SD. Contrasts are higher minus lower ability. Inside-zone comparisons contain 4,425 pairs across nine conditions; outside comparisons contain 13,235 pairs across 12. Secure-lead differences inside are positive in all nine conditions; weak-return differences outside are negative in 11 of 12. Matches chiefly compare middle- and lower-ability agents. Restricting to the top and bottom ability quartiles leaves 215 pairs from 60 trajectories and nine conditions. This does not represent all extreme-ability agents.

Pair differences are averaged within condition, then equally across supported conditions. Repeated agents/pairs are not independent replications. With one simulation per configuration, contrasts are descriptive; the two matched populations do not identify relative causal importance.

The within-agent analysis fits descriptive linear probability models with agent and condition-by-round fixed effects. The regressor $(200-r)/20$ expresses a 20-place rank improvement . Secure-lead and unlikely-catch-up coefficients are +5.80 and $-4.95$ percentage points per 20-place improvement. Rank, scores, and experiences update together, preventing causal isolation.

\subsection{Coverage and Sensitivity Checks}
Post-feedback substantive coverage is 90.70\% (D1), 50.34\% (D2), 29.62\% (D3), 59.93\% (D4), and approximately 0.10\% (D5); these rates differ from broad topic mentions in Section~\ref{sec:reasoning}.

Alternative rank-matching ability tolerances of 0.025 and 0.10 SD yield 24,120 and 33,819 pairs; the five-place ability-matching rank tolerance yields 15,820 pairs. The first-post-feedback rank comparison contains 5,942 pairs. Tightening the rank-comparison ability tolerance to 0.025 SD or widening it to 0.10 SD yields secure-lead differences of +8.54 and +7.69 points and unlikely-catch-up differences of $-13.53$ and $-13.96$ points. Using only the first decision after each feedback update gives +7.09 and $-13.75$ points. Tightening the ability-comparison rank tolerance to five places gives +2.06 for secure leads and +0.19 for unlikely catch-up. These checks preserve the main directions without establishing causality.

Removing 11 possible D2a false matches of ``margin'' within ``marginal returns,'' with pairs fixed, changes secure-lead contrasts from +7.956 to +7.963 points (rank) and +2.398 to +2.388 (ability). Main statistics retain original codes; the targeted check leaves the contrasts essentially unchanged.

\subsection{Supplementary Feedback Analysis}
Figure~\ref{fig:reasoning-feedback} supplements the main matched comparisons with within-agent responses around published feedback.

\subsection{Judgments Relative to the Reward Boundary}
\label{app:reasoning-boundary}
For $r-K\leq-10$, secure leads occur in 36.4\% of rationales; at $-9\leq r-K\leq0$, qualification threats reach 55.9\%. Just outside ($1\leq r-K\leq10$), attainable catch-up reaches 22.5\%; farther outside, unlikely catch-up reaches 24.7\%. Rates are averaged equally across configurations without ability matching. Thus, qualification threats track cutoff proximity rather than increase monotonically with worse rank.

\clearpage
\end{document}